\documentclass[prd,aps,preprint,tightenlines,superscriptaddress]{revtex4}
\usepackage{epsfig}
\usepackage{axodraw}
\usepackage{amsmath}
\begin{document}


\title{Collins-Soper Equation for the Energy Evolution of \\Transverse-Momentum and Spin Dependent Parton Distributions}
\author{Ahmad Idilbi}
\affiliation{Department of Physics, University of Maryland,
College Park, Maryland 20742, USA}
\author{Xiangdong Ji}
\affiliation{Department of Physics,
University of Maryland,
College Park, Maryland 20742, USA}
\author{Jian-Ping Ma}
\affiliation{Institute of Theoretical Physics,
Academia Sinica, Beijing, 100080, P. R. China}
\author{Feng Yuan}
\affiliation{Department of Physics,
University of Maryland,
College Park, Maryland 20742, USA}

\date{\today}
\vspace{0.5in}
\begin{abstract}
The hadron-energy evolution (Collins and Soper) equation for all
the leading-twist transverse-momentum and spin dependent parton
distributions is derived in the impact parameter space. Based on
this equation, we present a resummation formulas for the spin
dependent structure functions of the semi-inclusive deep inelastic
scattering.
\end{abstract}

\maketitle

\section{Introduction}
Semi-inclusive processes at low transverse momentum have attracted
considerable interest in recent years. These processes can provide
much information on aspects of perturbative and non-perturbative
quantum chromodynamics (QCD), and on the internal structure of the
nucleon in particular. Theoretical study of these processes began
with the classical work of Collins and Soper in which a nearly
back-to-back hadron pair is produced in $e^+e^-$ collision
\cite{ColSop81}. Applications to the Drell-Yan process and the
semi-inclusive deep-inelastic scattering (SIDIS) were made lately
in \cite{ColSopSte85,MenOlnSop96,NadStuYua00}. A factorization
theorem for the process was established \cite{ColSop81}, involving
a new class of non-perturbative hadronic observables depending on
the transverse-momentum of hadrons and/or partons: the
transverse-momentum dependent (TMD) fragmentation functions and
parton distributions. Based on the recent development of the gauge
invariant definition of the TMD parton distributions
\cite{{Col02},BelJiYua03}, the proof of the factorization has been
extended to the SIDIS and Drell-Yan processes at low transverse
momentum (one the order of $\Lambda_{QCD}$)
\cite{JiMaYu04,JiMaYu04p}, where the TMD parton distributions and
fragmentation functions as important ingredients were emphasized.

On the other hand, at low transverse momentum, $P_\perp\ll Q$,
where $Q$ represents some hard scale (e.g. the invariant mass of
the virtual photon in SIDIS), the standard perturbative QCD (pQCD)
calculations generate large logarithms $\alpha_s^n\ln^{2n}
Q^2/P_\perp^2$ in perturbation series. These large logarithms must
be resummed to make predictions reliable
\cite{ColSopSte85,DokDyaTro80,ParPet79}. According to the
factorization theorem for the semi-inclusive processes, these
large logarithms can be attributed to the hadron-energy dependence
of the TMD parton distributions and fragmentation functions, which
contains double logarithms \cite{ColSop81,JiMaYu04}. The
Collins-Soper equation is precisely the equation which governs the
energy dependence of the TMD parton distributions and the
fragmentation functions \cite{ColSop81}. The factorization proof
and an extensive study for the energy evolution equation for the
unpolarized quark distribution have been performed in
\cite{ColSop81}. The large logarithms mentioned above can be
resummed by solving this evolution equation
\cite{ColSop81,{ColSopSte85}}. In this paper, we will follow these
studies to analyze the evolution equations for the spin-dependent
quark distributions. We will show that the original Collins-Soper
evolution equation also applies to the spin-dependent
distributions. With these evolution equations, we can perform the
large logarithmic resummations for the spin-dependent structure
functions and asymmetries in the semi-inclusive DIS
\cite{JiMaYu04,JiMaYu04p}. Another important point of our
calculations is that we use Feynman gauge, and the TMD parton
distributions in this gauge are defined in such a way to guarantee
the gauge invariance \cite{{Col02},BelJiYua03}. In
\cite{ColSop81}, a specified gauge (axial gauge) was used, while
in \cite{Col89} the energy evolution for the Sudakov form factor
was calculated in Feyman gauge.

The rest of this paper is organized as follows. In Sec.II, we
briefly review the factorization of the energy evolution equation
for the TMD quark distribution following \cite{ColSop81,Col89},
and point out that this factorization works also for the
spin-dependent distributions. The soft and hard parts in the
factorization are defined. In Sec.III, the evolution equation for
the unpolarized quark distribution at one-loop order is rederived.
In Sec.IV, we show the results for all the leading-twist TMD quark
distributions. In Sec.V, with these evolution equations, we show
how to resum the large logarithms. We conclude in Sec.VI.

\section{Factorization of the Energy Derivative for the TMD quark distributions}

In the non-singular gauge (e.g. Feynman gauge), the TMD quark
distributions can be defined through the following matrix:
\cite{Col89,Col02,clc},
\begin{eqnarray}
       {\cal M}(x, k_\perp, \mu, x\zeta,\rho)
        &=& p^+\int
        \frac{d\xi^-}{2\pi}e^{-ix\xi^-P^+}\int
        \frac{d^2\vec{b}_\perp}{(2\pi)^2} e^{i\vec{b}_\perp\cdot
        \vec{k}_\perp} \nonumber \\
   &&    \times \frac{\left\langle PS\left|\overline{\psi}_q(\xi^-,0,\vec{b}_\perp){\cal
        L}^\dagger_{v}(\infty;\xi^-,0,\vec{b}_\perp)  {\cal
        L}_{v}(\infty;0)
        \psi_q(0)\right|PS\right\rangle}{S(\vec{b}_\perp, \mu^2, \rho) } \ ,
        \label{tmdpd}
\end{eqnarray}
where $\psi_q$ is the quark field with the Dirac- and color
indices implicit. The parent hadron has a momentum $P^\mu$ along
the $z$-direction, and is polarized with a spin vector $S^\mu$. In
the following, we use the light-cone coordinates $k^\pm = (k^0\pm
k^3)/\sqrt{2}$, and write any four-vector $k^\mu$ in the form of
$(k^-,\vec{k})= (k^-, k^+, \vec{k}_\perp)$, where $\vec{k}_\perp$
represents two perpendicular components $(k^x,k^y)$. The
light-like vector $p$ is chosen to be
$p^\mu=\Lambda(0,1,0_\perp)$, and $v^\mu$ is a time-like
dimensionless ($v^2>0$) four-vector with zero transverse
components $(v^-,v^+,\vec{0})$. We choose $v^-\gg v^+$ so that
$v^\mu$ is very close to the light-like vector $n^\mu =
(1,0,0_\perp)/2\Lambda $. The variable $\zeta^2$ is essentially
the energy of the hadron, $\zeta^2=(2P\cdot
v)^2/v^2=2(P^+)^2v^-/v^+$. ${\cal L}_v$ is a gauge link along
$v^\mu$,
\begin{equation}
    {\cal L}_{v}(\infty;\xi) = \exp\left(-ig\int^{\infty}_0 d\lambda v\cdot A(\lambda
    v +\xi)\right) \ .
\end{equation}
Here a non-light-like gauge link is introduced to regulate the
light-cone singularity associated with the plus component of the
gluon momentum $l^+\rightarrow 0$ \cite{ColSop81,Col89}. We avoid
the use of the singular gauge (e.g. the light-cone gauge), because
in such a gauge it is well known that the gauge potential does not
vanish at infinity, and therefore a gauge link at infinity will be
needed to ensure gauge invariance \cite{BelJiYua03}.

In the above definition, we have divided by a soft factor which is
defined as \cite{JiMaYu04}:
\begin{equation}
   S(\vec{b}_\perp, \mu^2, \rho) =\frac{1}{N_c} \langle 0|
   {\cal L}^\dagger_{\tilde vil}( \vec{b}_\perp, -\infty)
   {\cal L}^\dagger_{vlj}(\infty;\vec{b}_\perp)
   {\cal L}_{vjk}(\infty;0)
    {\cal L}_{\tilde v ki}(0;-\infty) |0\rangle\ ,
\label{soft}
\end{equation}
where $\tilde v$ is another off-light-cone vector with $\tilde
v^+\gg \tilde v^-$ and $\tilde v_\perp=0$. The parameter $\rho$ is
defined as $\rho^2=v^-\tilde v^+/v^+\tilde v^-$. This soft factor
will also appear in the factorization theorem for the
semi-inclusive processes at low $P_\perp$ \cite{JiMaYu04}.

The above definition of the matrix ${\cal M}$ is for deep
inelastic scattering. For the Drell-Yan process, we need a
different form where the gauge links are along the backward
direction to $-\infty$ \cite{Col02,BelJiYua03}, and the soft
factor has to be modified as well \cite{JiMaYu04p}. In the
following discussions, we will restrict ourselves to those for
SIDIS, although the same can be done for the distributions in the
Drell-Yan process.

The leading-twist expansion of the matrix ${\cal M}$ contains
eight distributions \cite{MulTan96,BoeMul98},
\begin{eqnarray}
{\cal M}&=&\frac{1}{2}\left[q(x,k_\perp)\not\! p+\frac{1}{M}\delta
q(x,k_\perp)\sigma^{\mu\nu}k_\mu p_\nu+\Delta q_L(x,k_\perp) \lambda \gamma_5\not\! p  \right.\label{tmdpar}\\
&& +\frac{1}{M}\Delta q_T(x,k_\perp)\gamma_5\not\!
p(\vec{k_\perp}\cdot \vec{S}_\perp)+\frac{1}{M}\delta
q_L(x,k_\perp)\lambda i\sigma_{\mu\nu}\gamma_5 p^\mu k_\perp^\nu
+\delta q_T(x, k_\perp)i\sigma_{\mu\nu}\gamma_5 p^\mu S_\perp^\nu
\nonumber\\
&&\left.+\frac{1}{M^2}\delta
q_{T'}(x,k_\perp)i\sigma_{\mu\nu}\gamma_5 p^\mu
\left(\vec{k}_\perp\cdot\vec{S}_\perp
k_\perp^\nu-\frac{1}{2}\vec{k}_\perp^2S_\perp^\nu\right)
+\frac{1}{M}q_T(x,k_\perp)\epsilon^{\mu\nu\alpha\beta}\gamma_\mu
p_\nu k_\alpha S_\beta \right]\ , \nonumber
\end{eqnarray}
where $M$ is the nucleon mass. We have omitted the arguments
$\zeta$, $\mu$ and $\rho$ in the parton distributions on the
right-hand side. The polarization vector $S^\mu$ has been
decomposed into a longitudinal component $S_L^\mu$ and a
transverse one $S_\perp^\mu$, and $\lambda$ is the helicity. The
notations for the distributions follow Ref. \cite{JiMaYu03}.

The energy evolution of the TMD parton distribution is governed by
the Collins-Soper equation \cite{ColSop81},
\begin{equation}
\zeta\frac{\partial}{\partial
\zeta}f(x,b,\mu,x\zeta,\rho)=\left(K(b,\mu,\rho)+G(x\zeta,\mu,\rho)\right)f(x,b,\mu,x\zeta,\rho)
\ , \label{e5}
\end{equation}
where $K$ and $G$ depend on soft and hard physics, respectively.
In the following we will show that this equation is in fact valid
for all the leading-twist quark distributions in the impact
parameter space. It is important to note that the above
factorization of soft and hard parts for the energy derivative is
true only for the leading power ($1/\zeta$) contribution, and any
higher power corrections have been neglected
\cite{ColSop81,Col89}. So, the evolution equations like
Eq.~(\ref{e5}) and here after are only valid in the leading power
of $1/\zeta$. Another point we want to point out is the $\rho$
dependence in $K$ and $G$, which was absent in
\cite{ColSop81,Col89}. The $\rho$ dependence comes from our
definition of $K$ in Eq.~(\ref{softk}) below to avoid possible
light-cone singularity in higher loop calculations. However, the
$\rho$ dependence of $K$ and $G$ cancels out, and energy
derivative of the TMD parton distribution does not depend on
$\rho$, (see detailed discussion below).

Since there is no energy dependence in the soft factor $S$, the
only source for energy dependence comes from the numerator in the
matrix ${\cal M}$ in Eq.~(\ref{tmdpd}). We call the numerator the
un-subtracted TMD parton distribution ${\cal Q}(x,k_\perp,x\zeta)$
as in \cite{JiMaYu04,JiMaYu04p}. So, the Collins-Soper equation
for any un-subtracted distribution reads
\begin{equation}
\zeta\frac{\partial}{\partial \zeta}{\cal
F}(x,b,\mu,x\zeta)=\left(K(b,\mu,\rho)+G(x\zeta,\mu,\rho)\right){\cal
F}(x,b,\mu,x\zeta) \ . \label{cse}
\end{equation}
where $K$ and $G$ depend on soft and hard physics, respectively,
as we shall explain. Since any un-subtracted distribution does not
depend on the soft part, then there is no $\rho$ dependence in
${\cal F}$. This means that sum
$\left(K(b,\mu,\rho)+G(x\zeta,\mu,\rho)\right)$ is
$\rho$-independent while each term could be $\rho$-dependent.

The different TMD quark distributions can be obtained from Eq. (4)
by making appropriate spin projections. For example, the
unpolarized quark distribution is related to,
\begin{eqnarray}
       {\cal Q}(x, k_\perp, \mu, x\zeta)
        &=& \frac{1}{2}\int
        \frac{d\xi^-d^2\vec{b}_\perp}{(2\pi)^3}e^{-ix\xi^-P^++i\vec{b}_\perp\cdot
        \vec{k}_\perp}  \left\langle P\left|\overline{\psi}_q(\xi^-,\vec{b}_\perp){\cal
        L}^\dagger_{v} \gamma^+ {\cal
        L}_{v}        \psi_q(0)\right|P\right\rangle\ .
        \label{tmdpd1}
\end{eqnarray}
Since $\zeta^2=(2v\cdot P)^2/v^2$, using the chain rule, the
derivative on $\zeta$ can be related to the derivative on $v$,
\begin{equation}
\zeta\frac{\partial}{\partial \zeta}=\delta
v^\alpha\frac{\partial}{\partial v^\alpha} \ ,
\end{equation}
where $\delta v$ is another dimensionless vector: $\delta
v^-=v^-$, $\delta v^+=-v^+$, and $\delta v_\perp=0$. So that, we
have $\delta v^2=-v^2<0$ and $\delta v\cdot v=0$. From
Eq.~(\ref{tmdpd1}), we see that the only dependence of $v$ comes
from the gauge link ${\cal L}_v$. So, we have the following
differential equation for the un-subtracted TMD parton
distribution,
\begin{eqnarray}
&&\zeta\frac{\partial}{\partial \zeta}{\cal Q}(x,k_\perp,v)=\delta
v^\alpha\frac{\partial}{\partial v^\alpha}{\cal
Q}(x,k_\perp,v)\nonumber\\&&=\frac{1}{2}\int
        \frac{d\xi^-d^2\vec{\xi}_\perp}{(2\pi)^3}e^{-ix\xi^-P^++i\vec{\xi}_\perp\cdot
        \vec{k}_\perp}  \\
        &&\times\left\{\left\langle P\left|\overline{\psi}_q(\xi){\cal
        L}^\dagger_{v} \gamma^+ (-ig)\int_0^\infty d\lambda\left[\delta v\cdot A(\lambda v)
        +\lambda \delta v\cdot \partial A(\lambda v)\cdot  v \right]  {\cal
        L}_{v}
        \psi_q(0)\right|P\right\rangle\right.\nonumber\\
&&+\left.\left\langle P\left|\overline{\psi}_q(\xi){\cal
        L}^\dagger_{v} (ig)\int_0^\infty d\lambda\left[\delta v\cdot A(\lambda v+\xi)
        +\lambda \delta v\cdot \partial A(\lambda v+\xi)\cdot v \right]\gamma^+ {\cal
        L}_{v}
        \psi_q(0)\right|P\right\rangle\right\} \ . \nonumber
\end{eqnarray}
The relevant Feynman rules for the eikonal vertex  from the gauge
link contribution are shown in Fig.~1 for the TMD parton
distributions (a), and the derivative
$\zeta\frac{\partial}{\partial \zeta}$ (b).

\begin{figure}
\begin{center}
\begin{picture}(200,100)(0,0) \SetOffset(0,50)
\SetWidth{0.7} \Line(0,46)(22,46)\Line(0,44)(22,44)
\Gluon(22,10)(22,46){-2}{6}\Line(9,42)(12,45)\Line(9,48)(12,45)\Line(20,31)(23,28)\Line(26,31)(23,28)
\Text(35,28)[l]{\large $=\frac{\large i}{\large v\cdot
k+i\epsilon}(-ig)v^\mu=\frac{\large g}{\large v\cdot
k+i\epsilon}v^\mu$}

\SetOffset(0,0) \SetWidth{0.7}
\Line(0,46)(22,46)\Line(0,44)(22,44)
\Gluon(22,10)(22,46){-2}{6}\Line(9,42)(12,45)\Line(9,48)(12,45)\Line(20,31)(23,28)\Line(26,31)(23,28)
\Text(35,28)[l]{\large $=g\frac{v\cdot k \delta v^\mu-\delta
v\cdot k v^\mu}{\large (v\cdot k+i\epsilon)^2}$}\SetColor{Blue}
\Line(19,42)(25,48)\Line(19,48)(25,42)\SetColor{Blue}
\end{picture} \end{center}

\caption{\it Feynman rules for the eikonal vertex in the TMD
parton distribution (a) and its derivative (b).}
\end{figure}

Similar to the structure functions for the semi-inclusive DIS
discussed in \cite{JiMaYu04, JiMaYu04p}, the derivative
$\zeta\frac{\partial}{\partial \zeta}$ on the un-subtracted
distribution ${\cal Q}$ receives contributions from three
different regions of the gluon momentum: soft, hard, and collinear
contributions, respectively.

First, let us examine the collinear contribution to the
derivative. From the above, the special vertex reads,
\begin{equation}
g\frac{v\cdot k\delta v^\mu-\delta v\cdot k v^\mu}{(v\cdot
k+i\epsilon)^2}\ .
\end{equation}
If the gluon momentum $k$ is in the collinear region, i.e.,
$k\propto P$, we have $k^+\sim Q$, $k^-\sim \lambda Q$,
$k_\perp\sim \sqrt{\lambda} Q$, where $\lambda$ is a small
parameter. For such momentum, the above vertex will lead to
$$
\sim g \frac{k^+v^-}{(v\cdot k+i\epsilon)^2}(\delta v-v)^\mu\ ,
$$
and their contribution will be suppressed if we contract the above
vertex with the collinear momentum in the jet part of the TMD
parton distributions.

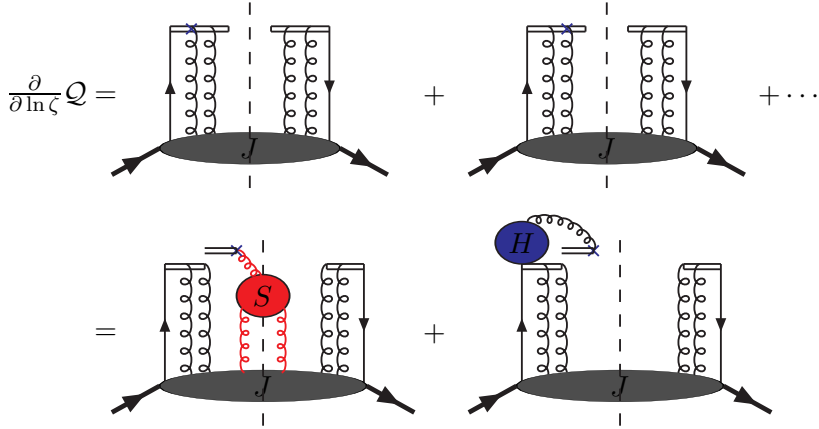
\begin{figure}
\begin{center}
\begin{picture}(330,170)(0,0)

\SetOffset(50,100) \SetWidth{0.7}
\ArrowLine(10,10)(10,56)\ArrowLine(70,56)(70,10)
\Line(10,56)(32,56)\Line(10,54)(32,54)\Line(32,54)(32,56)
\SetColor{Blue}\Line(16,53)(20,57)\Line(16,57)(20,53)\SetColor{Black}
\Line(70,56)(48,56)\Line(70,54)(48,54)\Line(48,54)(48,56)
\Gluon(25,10)(25,56){-2}{6}\Gluon(18,10)(18,56){-2}{6}
\Gluon(55,10)(55,56){2}{6}\Gluon(63,10)(63,56){2}{6}
\GOval(40,10)(34,6)(90){0.3}\Text(40,10)[c]{$J$}\DashLine(40,65)(40,-5){5}
\SetWidth{1.8}\ArrowLine(-12,0)(6,10) \ArrowLine(74,10)(92,0)
\Text(110,30)[c]{$+$} \Text(-10,30)[r]{
$\frac{\partial}{\partial\ln\zeta}{\cal Q}=$}

\SetOffset(185,100)\SetWidth{0.7}
\ArrowLine(10,10)(10,56)\ArrowLine(70,56)(70,10)
\Line(10,56)(32,56)\Line(10,54)(32,54)\Line(32,54)(32,56)
\Line(70,56)(48,56)\Line(70,54)(48,54)\Line(48,54)(48,56)
\SetColor{Blue}\Line(23,53)(27,57)\Line(23,57)(27,53)\SetColor{Black}
\Gluon(25,10)(25,56){-2}{6}\Gluon(18,10)(18,56){-2}{6}
\Gluon(55,10)(55,56){2}{6}\Gluon(63,10)(63,56){2}{6}
\GOval(40,10)(34,6)(90){0.3}\Text(40,10)[c]{$J$}
\DashLine(40,65)(40,-5){5} \SetWidth{1.8}\ArrowLine(-12,0)(6,10)
\ArrowLine(74,10)(92,0) \Text(110,30)[c]{$+\cdots$}

\SetOffset(50,10) \SetWidth{0.7}
\ArrowLine(8,10)(8,56)\ArrowLine(83,56)(83,10)
\Line(8,56)(23,56)\Line(8,54)(23,54)\Line(23,54)(23,56)
\Line(83,56)(69,56)\Line(83,54)(69,54)\Line(69,54)(69,56)
\Gluon(23,10)(23,56){-2}{6}\Gluon(16,10)(16,56){-2}{6}
\Gluon(69,10)(69,56){2}{6}\Gluon(75,10)(75,56){2}{6}
\GOval(45,10)(39,6)(90){0.3}\Text(45,10)[c]{$J$}\DashLine(45,65)(45,-5){5}
\COval(45,44)(10,8)(90){Black}{Red}\Text(45,44)[c]{$S$}\SetColor{Red}
\Gluon(38,40)(38,15){-1.5}{4}\Gluon(52,40)(52,15){1.5}{4}
\Gluon(35,61)(44,50){1.5}{3} \SetColor{Blue}
\Line(33,59)(37,63)\Line(33,63)(37,59) \SetColor{Black}
\Line(23,60)(35,60)\Line(23,62)(35,62)\Line(35,60)(35,62)

\SetWidth{1.8}\ArrowLine(-12,0)(6,10) \ArrowLine(84,10)(102,0)
\Text(110,30)[c]{$+$} \Text(-10,30)[r]{ $=$}

\SetOffset(185,10) \SetWidth{0.7}
\ArrowLine(8,10)(8,56)\ArrowLine(83,56)(83,10)
\Line(8,56)(23,56)\Line(8,54)(23,54)\Line(23,54)(23,56)
\Line(83,56)(69,56)\Line(83,54)(69,54)\Line(69,54)(69,56)
\Gluon(23,10)(23,56){-2}{6}\Gluon(16,10)(16,56){-2}{6}
\Gluon(69,10)(69,56){2}{6}\Gluon(75,10)(75,56){2}{6}
\GOval(45,10)(39,6)(90){0.3}\Text(45,10)[c]{$J$}\DashLine(45,65)(45,-5){5}
\COval(8,64)(10,8)(90){Black}{Blue}\Text(8,64)[c]{$H$}
\SetColor{Blue} \Line(33,59)(37,63)\Line(33,63)(37,59)
\SetColor{Black}\GlueArc(17,53)(20,25,110){1.5}{6}
\Line(23,60)(35,60)\Line(23,62)(35,62)\Line(35,60)(35,62)

\SetWidth{1.8}\ArrowLine(-12,0)(6,10) \ArrowLine(84,10)(102,0)

\end{picture} \end{center} \caption{\it
Factorization for the derivative $\partial/\partial \ln\zeta$ on
the un-subtracted TMD parton distribution.}
\end{figure}

Thus, the derivative $\zeta\frac{\partial}{\partial \zeta}$ on the
parton distribution will receive contributions only from the soft
and hard regions of the gluon momentum. A detailed analysis of
these contributions leads to a factorization of the derivative of
the TMD distribution as illustrated in the second line in Fig.~2
\cite{ColSop81,Col89}. This can be represented by the following
differential equation,
\begin{equation}
\zeta\frac{\partial}{\partial \zeta}{\cal
Q}(x,k_\perp,x\zeta)=\int \left[K +G\right ]\otimes {\cal
Q}(x,k_\perp,x\zeta) \ ,
\end{equation}
where the soft part is called $K$, and hard part $G$; and the
label $\otimes$ means the momentum space convolution. Again, we
emphasize that the above factorization is only valid for the
leading power contribution. After applying a Fourier
transformation to the coordinate space, the above differential
equation reads,
\begin{equation}
\zeta\frac{\partial}{\partial \zeta}{\cal Q}(x,b,x\zeta)=
\left[K(b,\mu,\rho) +G(x\zeta,\mu,\rho)\right ]\times {\cal
Q}(x,b,x\zeta) \ ,
\end{equation}
where the convolution in momentum space becomes products in the
impact parameter $b$-space. The soft-part $K$ depends on $b$,
while the hard part $G$ depends on hard scale $\zeta$; and both of
them depend on the renormalization scale $\mu$ and $\rho$, but the
sum does not. The above equation is valid for any value of $b$. If
$b$ is small as $1/b\gg \Lambda_{QCD}$, we can further have a
factorization for the TMD parton distribution which depends on the
integrated parton distributions, and then we can get another form
for the evolution equation with an extra term \cite{ColSop81}.

The soft part can be calculated from the Feynman diagrams by using
the Grammer-Yennie approximation \cite{GraYen73}. In
\cite{JiMaYu04}, we demonstrated how to get the soft contribution
in the TMD parton distribution. Here, we follow the same
procedure, and define the soft part for the derivative of the TMD
quark distributions by the following matrix element,
\begin{eqnarray}
\label{softk}
 K(b,\mu,\rho)&=&\left\langle 0\left|{\cal
        L}^\dagger_{\tilde v}(\vec{b}_\perp,-\infty){\cal
        L}^\dagger_{v}(\infty,\vec{b}_\perp) \left\{(-ig)\int_0^\infty d\lambda\left[\delta v\cdot A(\lambda v)
        +\lambda \delta v\cdot \partial A(\lambda v)\cdot v\nonumber\right] \right.\right.\right.\\
        &&\left.\left.\left.+(ig)\int_0^\infty d\lambda\left[\delta v\cdot A(\lambda v+b_\perp)
        +\lambda \delta v\cdot \partial A(\lambda v+b_\perp)\cdot v \right]\right\}{\cal
        L}_{v}(\infty,0)
        {\cal
        L}_{\tilde v}(0,-\infty)\right|0\right\rangle
       \nonumber \\
       &&\left/\left\langle 0\left|{\cal
        L}^\dagger_{\tilde v}(\vec{b}_\perp,-\infty){\cal
        L}^\dagger_{v}(\infty,\vec{b}_\perp) {\cal
        L}_{v}(\infty,0)
        {\cal
        L}_{\tilde v}(0,-\infty)\right|0\right\rangle \right. \
        ,
\end{eqnarray}
where we have introduced the off-light-cone vector $\tilde v$
defined above to regulate possible light-cone singularities
\cite{JiMaYu04}. Notice that this definition is different from
that in \cite{Col89} where a light-cone vector $p=(0,1,0_\perp)$
was used instead of our off-light-cone vector $\tilde v$.
Introducing $\tilde v$ leads to the $\rho$ dependence of $K$.
Since the leading order (one-loop) calculation has no light-cone
singularity, there is no difference between using $\tilde v$ and
$p$ at this order, and the result does not depend on $\rho$
\cite{Col89}. However, at higher-loop, there might be light-cone
singularity associated with the gluon momentum $l^+\to 0$. We note
that so far there is no explicit calculation of the soft factor
$K$ beyond one-loop order in non-abelian gauge theory (a QED
result has been given in \cite{Col89}) . It is not clear at
present that we can take the light-cone limit for higher loop
calculations. So, it is necessary to include the off-light-cone
vector $\tilde v$ in the formal definition. If there is no
light-cone singularity, we can take the light-cone limit ($\tilde
v\to p$ and $\rho\to \infty$). Like in the factorization of SIDIS
structure function \cite{JiMaYu04}, $\rho$ is just a parameter
which separates the hard and soft physics, and it does not affect
any prediction power in the resummation formalism. Normally, we
should take $\rho\gg 1$, while in practice we can choose $\rho$
between 3 and 10 if there is any $\rho$ dependence in $K$ to avoid
the large logarithms associate with $\rho$.

Comparing Eq.~(\ref{softk}) with Eq.~(\ref{soft}), we find that
the soft factor in the derivative can be related to that in the
parton distribution,
\begin{equation}
K(b,\mu,\rho)=\frac{1}{S(b,\mu,\rho)}\delta
v^\alpha\frac{\partial}{\partial
v^\alpha}S(b,\mu,\rho)=\frac{1}{S(b,\mu,\rho)}\rho\frac{\partial}{\partial
\rho}S(b,\mu,\rho) \ . \label{softsk}
\end{equation}
Our one-loop result below verifies this relation. Since the
distribution is $\rho$-independent, the $\rho$ dependence in $K$
must cancel the $\rho$ dependence in $G(x\zeta,\mu,\rho)$.

The renormalization scale dependence of the soft-factor $K$ is
determined by the cusp anomalous dimension \cite{KorRad92},
\begin{equation}
\mu\frac{d}{d\mu}K=-\gamma_K \ ,
\end{equation}
which is a series in $\alpha_s$ and free of infrared
singularities. The hard part $G$ can be calculated through a
systematic subtraction, as will be illustrated by the one-loop
example in the next section. From the definition, it is obvious
that the soft factor is spin-independent.

The above analysis of the factorization of energy derivative and
the definitions of the hard and soft parts can be extended to the
spin-dependent TMD quark distributions, because they all come from
the same matrix ${\cal M}$ Eq.~(1) with different spin
projections, and the arguments supporting the factorization can be
generalized to all the leading-twist distributions. We will
explicitly show this in more detail in the following one-loop
calculations, and give general argument for all orders.

\section{Re-derivation of Collins-Soper Equation for Unpolarized Quark Distribution}

In this section, we demonstrate how to calculate, in the present
formulation, the Collins-Soper evolution kernels $K$ and $G$ for
the unpolarized quark distribution ${\cal Q}$ at one-loop order.
In next section, we will discuss the spin dependence and present
the evolution equations for all the leading-twist TMD quark
distributions.

\begin{figure}
\begin{center} \begin{picture}(400,90)(0,0)
\SetWidth{1.0}

\SetOffset(0,15)
\Line(0,0)(0,30)\Line(3,0)(3,30)\Line(0,30)(20,60)\Line(3,30)(23,60)
\Line(65,30)(65,0)\Line(62,30)(62,0)\Line(65,30)(45,60)\Line(62,30)(42,60)
\GlueArc(12,45)(8,235,418){1.5}{5}\SetColor{Blue}\Line(12,52)(20,52)\Line(16,48)(16,56)
\SetColor{Black}
\DashLine(32,65)(32,-5){5}\Text(83,30)[c]{+}\Text(32,-15)[c]{(a)}

\SetOffset(100,15)
\Line(0,30)(0,0)\Line(3,30)(3,0)\Line(0,30)(20,60)\Line(3,30)(23,60)
\Line(65,30)(65,0)\Line(62,30)(62,0)\Line(65,30)(45,60)\Line(62,30)(42,60)
\SetColor{Blue}\Line(7,44)(15,44)\Line(11,40)(11,48)
\SetColor{Black} \GlueArc(5,30)(14,255,428){2}{8}
\DashLine(32,65)(32,-5){5}\Text(32,-15)[c]{(b)}\Text(83,30)[c]{+}

\SetOffset(200,15)
\Line(0,30)(0,0)\Line(3,30)(3,0)\Line(0,30)(20,60)\Line(3,30)(23,60)
\Line(65,30)(65,0)\Line(62,30)(62,0)\Line(65,30)(45,60)\Line(62,30)(42,60)
\Gluon(11,44)(54,44){-2}{7}
\SetColor{Blue}\Line(7,44)(15,44)\Line(11,40)(11,48)
\SetColor{Black}
\DashLine(32,65)(32,-5){5}\Text(32,-15)[c]{(c)}\Text(83,30)[c]{+}

\SetOffset(300,15)
\Line(0,30)(0,0)\Line(3,30)(3,0)\Line(0,30)(20,60)\Line(3,30)(23,60)
\Line(65,30)(65,0)\Line(62,30)(62,0)\Line(65,30)(45,60)\Line(62,30)(42,60)
\Gluon(11,44)(65,15){-2}{9}
\SetColor{Blue}\Line(7,44)(15,44)\Line(11,40)(11,48)
\SetColor{Black} \DashLine(32,65)(32,-5){5}\Text(32,-15)[c]{(d)}

\end{picture}
\end{center}
\caption{\it One-loop diagrams for the soft part $K$. The mirror
diagrams are not shown, but included in the result.}
\end{figure}
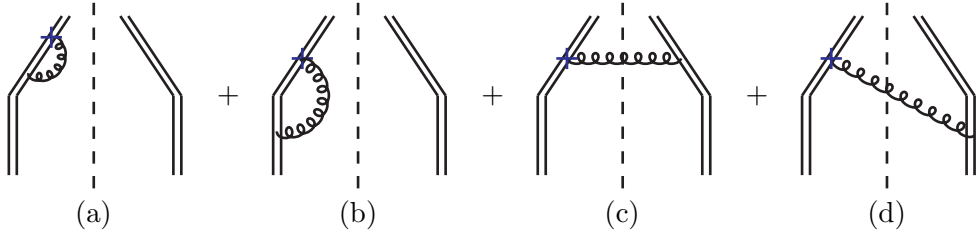

We first calculate the leading contribution to the soft part $K$,
linearly proportional to the strong coupling constant $\alpha_s$.
The relevant diagrams are shown in Fig.~3. Fig.~3(a) vanishes
because $\delta v\cdot v=0$, and 3(c) vanishes for the same reason
and for being sub-leading order in $1/\zeta^2$. The contribution
from Fig.~3(b) is
\begin{equation}
K(b,\mu,\rho)|_{\rm fig.3(b)}=-\frac{\alpha_s
C_F}{\pi}\ln\frac{\mu^2 }{\lambda^2} \ ,\label{b3}
\end{equation}
where we have included a factor of two to account for the two
vertex correction diagrams. $\lambda$ is the gluon mass,
introduced to regulate the infrared singularity for individual
diagrams. However, the sum of all contributions is free of the
infrared divergence. Fig.~3(d), together with its mirror diagram,
contributes
\begin{equation}
K(b,\mu,\rho)|_{\rm fig.3(d)}=\frac{\alpha_s
C_F}{\pi}\left[\ln\frac{4 }{b^2\lambda^2}-2\gamma_E \right]\ .
\label{d3}
\end{equation}
Summing up, we get,
\begin{equation}
K(b,\mu,\rho)=-\frac{\alpha_s C_F}{\pi}\left[\ln\frac{\mu^2
b^2}{4}+2\gamma_E\right] \ ,
\end{equation}
which agrees with the previous calculations
\cite{ColSop81,Col89,Li97}. Moreover, up to first order it has no
dependence on $\rho$, which means that we can take the light-cone
limit for $\tilde v$ (i.e., let $\tilde v=p$) at this order. We
remark here that higher order calculation of $K(b,\mu,\rho)$ may
show explicit $\rho$ dependence. Comparing this result with the
soft factor $S(b,\mu,\rho)$ at one-loop order calculated in
\cite{JiMaYu04}, Eq.~(\ref{softsk}) is clearly satisfied. The
one-loop cusp anomalous dimension is,
\begin{equation}
\mu\frac{\partial}{\partial
\mu}K(b,\mu)=-\gamma_K=-2\frac{\alpha_sC_F}{\pi} \ ,
\end{equation}
which is well known. As we stated in the previous section, the
soft part is spin independent, hence the above result for $K$ is
the same for all the leading order TMD quark distributions.

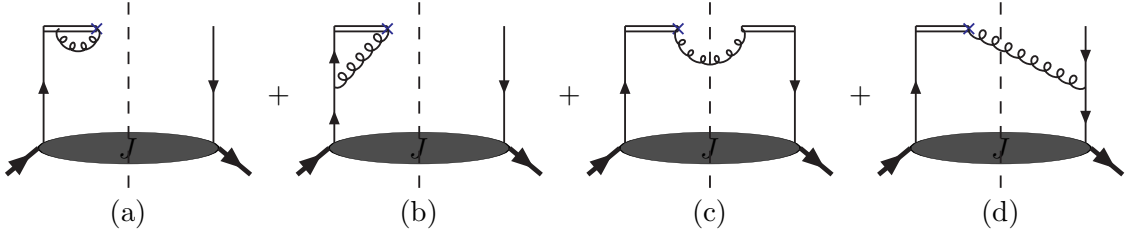
\begin{figure}
\begin{center}
\begin{picture}(450,70)(0,0)

\SetOffset(10,10) \SetWidth{0.7}
\ArrowLine(8,10)(8,56)\ArrowLine(72,56)(72,10)
\Line(8,56)(28,56)\Line(8,54)(28,54)\Line(28,54)(28,56)
\SetColor{Blue}\Line(26,53)(30,57)\Line(26,57)(30,53)\SetColor{Black}
\GlueArc(21,55)(7,180,360){1.5}{4}
\GOval(40,10)(34,6)(90){0.3}\Text(40,10)[c]{$J$}\DashLine(40,65)(40,-5){5}
\SetWidth{1.8}\ArrowLine(-6,0)(6,10) \ArrowLine(74,10)(86,0)
\Text(97,30)[c]{$+$}\Text(40,-15)[c]{(a)}

\SetOffset(120,10) \SetWidth{0.7}
\ArrowLine(8,10)(8,33)\ArrowLine(8,33)(8,56)\ArrowLine(72,56)(72,10)
\Line(8,56)(28,56)\Line(8,54)(28,54)\Line(28,54)(28,56)
\SetColor{Blue}\Line(26,53)(30,57)\Line(26,57)(30,53)\SetColor{Black}
\Gluon(8,33)(28,55){-2}{5}
\GOval(40,10)(34,6)(90){0.3}\Text(40,10)[c]{$J$}\DashLine(40,65)(40,-5){5}
\SetWidth{1.8}\ArrowLine(-6,0)(6,10) \ArrowLine(74,10)(86,0)
\Text(97,30)[c]{$+$}\Text(40,-15)[c]{(b)}

\SetOffset(230,10)\SetWidth{0.7}
\ArrowLine(8,10)(8,56)\ArrowLine(72,56)(72,10)
\Line(8,56)(28,56)\Line(8,54)(28,54)\Line(28,54)(28,56)
\SetColor{Blue}\Line(26,53)(30,57)\Line(26,57)(30,53)\SetColor{Black}
\Line(72,56)(52,56)\Line(72,54)(52,54)\Line(52,54)(52,56)
\GlueArc(40,55)(12,180,360){1.5}{6}
\GOval(40,10)(34,6)(90){0.3}\Text(40,10)[c]{$J$}\DashLine(40,65)(40,-5){5}
\SetWidth{1.8}\ArrowLine(-6,0)(6,10) \ArrowLine(74,10)(86,0)
\Text(97,30)[c]{$+$}\Text(40,-15)[c]{(c)}

\SetOffset(340,10)\SetWidth{0.7}
\ArrowLine(8,10)(8,56)\ArrowLine(72,56)(72,33)\ArrowLine(72,33)(72,10)
\Line(8,56)(28,56)\Line(8,54)(28,54)\Line(28,54)(28,56)
\SetColor{Blue}\Line(26,53)(30,57)\Line(26,57)(30,53)\SetColor{Black}
\Gluon(28,55)(72,33){-2}{8}
\GOval(40,10)(34,6)(90){0.3}\Text(40,10)[c]{$J$}\DashLine(40,65)(40,-5){5}
\SetWidth{1.8}\ArrowLine(-6,0)(6,10) \ArrowLine(74,10)(86,0)
\Text(40,-15)[c]{(d)}

\end{picture} \end{center} \caption{\it
One-loop diagrams for the Collins-Soper evolution for the
un-subtracted TMD parton distributions.}
\end{figure}

We now calculate the complete one-loop contribution to the
right-hand side of the Collins-Soper equation, from which we will
subtract the above soft contribution to get the hard part $G$. All
the one-loop diagrams are shown in Fig.~4. The contributions from
Fig.~4(a) and (c) vanish because of the same reason as that for
Fig.~3(a) and (c).

The contribution from Fig.~4(b) in momentum space reads,
\begin{eqnarray}
\frac{\partial}{\partial\ln\zeta}{\cal Q}(x,k_\perp,x\zeta)|_{\rm
fig.4(b)}&=&\frac{\alpha_sC_F}{\pi}\left(1-\ln\frac{x^2\zeta^2}{\lambda^2}\right){\cal
Q}(x,k_\perp,x\zeta) \ ,
\end{eqnarray}
which is the same as that in the impact parameter $b$ space. The
explicit dependence on $\zeta$ and $\lambda$ indicates the
presence of both hard and soft contributions. Subtracting the soft
contribution in Eq.~(\ref{b3}), we get the hard part $G$ as,
\begin{equation}
G(x\zeta,\mu)=\frac{\alpha_sC_F}{\pi}\left(1-\ln\frac{x^2\zeta^2}{\mu^2}\right)\
,
\end{equation}
which depends on the hard scale $\zeta$ and the renormalization
scale $\mu$.

Fig.~4(d) is dominated by the contribution from the soft gluon
momentum region: $q^+\ll k^+$, where $q$ is the gluon momentum and
$k$ is the quark momentum. After the soft-gluon approximation, we
get
\begin{eqnarray}
\frac{\partial}{\partial\ln\zeta}{\cal Q}(x,k_\perp,x\zeta)|_{\rm
fig.4(d)}&=&\frac{\alpha_sC_F}{2\pi^2}\int\frac{
d^2q_{\perp}}{{\vec q}_{\perp}^2+\lambda^2}{\cal
Q}(x,\vec{k}_\perp-\vec{q}_{\perp},x\zeta) \ .
\end{eqnarray}
Fourier-transforming to the $b$ space, we have
\begin{eqnarray}
\frac{\partial}{\partial\ln\zeta}{\cal Q}(x,b,x\zeta)|_{\rm
fig.4(d)}&=&\frac{\alpha_sC_F}{\pi}\left[\ln\frac{4}{b^2\lambda^2}-2\gamma_E\right]
{\cal Q}(x,b,x\zeta) \ ,
\end{eqnarray}
which can be reproduced by the soft factor $K$ from Eq.~(\ref{d3})
(Fig.~3(d)).

The above results show that the factorization is valid at one-loop
order with the sum of $K$ and $G$ reads,
\begin{equation}
K(b,\mu)+G(x\zeta,\mu)=-\frac{\alpha_sC_F}{\pi}\ln\frac{x^2\zeta^2b^2}{4}e^{2\gamma_E-1}
\ .
\end{equation}
This result agrees also with that in \cite{JiMaYu04}, where the
distribution itself was calculated to one-loop order.

\section{Collins-Soper Evolution Equations for Spin-Dependent TMD Distributions}

In this section, we study the energy evolution of the
spin-dependent TMD distributions. Except for the unpolarized
$q(x,k_\perp)$, all other leading-twist TMD quark distributions
depend on the polarization of either the initial hadron or the
probing quark. The polarized distributions can be obtained from
spin projections of the matrix ${\cal M}$ Eq.~(\ref{tmdpd}). There
are three leading-twist projections,
\begin{eqnarray}
\gamma^+&:&~~~q(x,k_\perp),~q_T(x,k_\perp) \ ;\nonumber\\
\gamma^+\gamma^5&:&~~~\Delta q_L(x,k_\perp),~\Delta
q_T(x,k_\perp)\
;\nonumber\\
\gamma^+\gamma^i\gamma^5&:&~~~\delta q_T(x,k_\perp),~\delta
q_L(x,k_\perp),~\delta q(x,k_\perp),~\delta q_T'(x,k_\perp) \ ,
\end{eqnarray}
corresponding to the unpolarized, longitudinally-polarized, and
transversely-polarized quark distributions, respectively.
Moreover, different distributions may have different $k_\perp$
dependence. For example, so-called $k_\perp$-even (under the
exchange $k_\perp \rightarrow -k_\perp$) quark distributions,
$q(x,k_\perp)$, $\Delta q_L(x,k_\perp)$, and $\delta
q_T(x,k_\perp)$, correspond to the unpolarized, helicity, and
transversity distributions, respectively \cite{JafJi91}. These
distributions survive after integrating over $k_\perp$. The other
five distributions are associated with the $k_\perp$-odd
structures, and vanish when $k_\perp$ are integrated over.

In the following, we use the Sivers function \cite{Siv90} as an
example to demonstrate the calculation of the energy evolution
kernel. The results for the other distributions can be obtained
similarly. The Sivers function results through $\gamma^+$
projection from the matrix ${\cal M}$ Eq.~(\ref{tmdpd}),
\begin{eqnarray}
       {\cal Q}_T(x, k_\perp, \mu, x\zeta)
        &=& \frac{1}{2\epsilon^{ij}S_ik_j}\int
        \frac{d\xi^-d^2\vec{b}_\perp}{(2\pi)^3}e^{-ix\xi^-P^++i\vec{b}_\perp\cdot
        \vec{k}_\perp} \nonumber\\
        &&\times \left.\left\langle PS_\perp\left|\overline{\psi}_q(\xi^-,\vec{b}_\perp){\cal
        L}^\dagger_{v} \gamma^+ {\cal
        L}_{v}        \psi_q(0)\right|PS_\perp\right\rangle\right|_{\rm spin~ dependent~ part}\ ,
        \label{sivers}
\end{eqnarray}
where the explicit transverse momentum and spin dependence has
been included. The Feynman diagrams for the energy derivative are
the same as those for the unpolarized distribution discussed in
last section; and Figs.~4(a) and (c) vanish as before. The
contribution from Fig.~4(d) reads,
\begin{eqnarray}
\frac{\partial}{\partial\ln\zeta}{\epsilon^{ij}S_ik_j\cal
Q}_T(x,k_\perp,x\zeta)|_{\rm
fig.4(d)}&=&\frac{\alpha_sC_F}{2\pi^2}\int\frac{
d^2q_{\perp}}{{\vec
q}_{\perp}^2+\lambda^2}{\epsilon^{ij}S_i(k-q)_j\cal
Q}_T(x,\vec{k}_\perp-\vec{q}_{\perp},x\zeta) \ ,\nonumber\\
\end{eqnarray}
where again we have made the soft approximation. To find the above
result, we have applied the $\gamma^+$ projection to the quark
matrix ${\cal M}$. Moreover, because the Sivers function is
spin-dependent and associated with a $k_\perp$-odd structure, only
such structure is isolated and kept. Fourier transforming to the
impact parameter space, we get,
\begin{equation}
\frac{\partial}{\partial\ln\zeta}\partial^i_b{\cal
Q}_T(x,b,x\zeta)|_{\rm
fig.4(d)}=\frac{\alpha_sC_F}{\pi}\left[\ln\frac{4}{b^2\lambda^2}-2\gamma_E\right]
\partial^i_b{\cal Q}_T(x,b,x\zeta) \ ,
\end{equation}
where $\partial^i_b=\partial/\partial b_i$ is a derivative on the
Sivers function. Apart from the explicit derivative, the above
contribution is the same as that for the unpolarized distribution
in the previous section.

The contribution from Fig.~4(b) is also the same as that for the
unpolarized quark distribution,
\begin{eqnarray}
\frac{\partial}{\partial\ln\zeta}{\cal
Q}_T(x,k_\perp,x\zeta)|_{\rm
fig.4(b)}&=&\frac{\alpha_sC_F}{\pi}\left(1-\ln\frac{x^2\zeta^2}{\lambda^2}\right){\cal
Q}_T(x,k_\perp,x\zeta) \ ,
\end{eqnarray}
and similar equation holds in the $b$-space. Combining the above
results, we have the entire energy evolution of the Sivers
function at one-loop order,
\begin{equation}
\zeta\frac{\partial}{\partial \zeta}\partial
^i_bq_T(x,b,\mu,x\zeta,\rho)=\left(K(b,\mu,\rho)+G(x\zeta,\mu,\rho)\right)\partial^i_bq_T(x,b,\mu,x\zeta,\rho)
\ ,
\end{equation}
where $K$ and $G$ are the same as those for the unpolarized
distribution.

The above analysis can be repeated for all other leading-twist
quark distributions, and the one-loop evolution kernels are found
again to be the same as those for the unpolarized quark
distribution. There are two important features supporting the
above finding. First, the leading-twist projection matrices
$\gamma^+$, $\gamma^+\gamma^5$, and $\gamma^+\gamma^i\gamma^5$
lead to an identical trace. Second, there is no mixing between the
different leading-twist distributions. This latter property,
however, does not hold for the evolution of the higher-twist
distributions, a topic beyond the scope of this paper.

In fact, we argue that, to all orders in perturbation theory, the
same evolution kernels determine the energy evolution of all the
leading-twist distributions. First of all, from the factorization
of the energy derivative discussed in the Section II, only soft
and hard regions contribute to the evolution. For the soft part,
there is no spin dependence, as is clear from its definition
Eq.~(\ref{softk}). Therefore, for any leading-twist distribution,
the soft part of the evolution kernel is the same. Second, the
contribution from the hard part is also spin-independent, because
the hard contribution is calculable in perturbative QCD and the
perturbative processes for massless quarks conserve helicity. Any
spin projection will lead to the same Dirac algebra if there is no
mixing between different distributions.

To summarize, the Collins-Soper evolution kernel has no spin
dependence for the leading-twist TMD quark distributions. For
$k_\perp$-even istributions, the following evolution equation,
\begin{equation}
\zeta\frac{\partial}{\partial
\zeta}f(x,b,\mu,x\zeta,\rho)=\left(K(b,\mu,\rho)+G(x\zeta,\mu,\rho)\right)f(x,b,\mu,x\zeta,\rho)
\ ,
\end{equation}
holds for $f=q$, $\Delta q_L$, and $\delta q_T$. For $k_\perp$-odd
quark distributions, we have
\begin{equation}
\zeta\frac{\partial}{\partial \zeta}\partial
^i_bf(x,b,\mu,x\zeta,\rho)=\left(K(b,\mu,\rho)+G(x\zeta,\mu,\rho)\right)\partial^i_bf(x,b,\mu,x\zeta,\rho)
\ ,
\end{equation}
which works for $f=q_T$, $\Delta q_T$, $\delta q$, and $\delta
q_L$. Finally, for $\delta q_T'$,
\begin{eqnarray}
&&\zeta\frac{\partial}{\partial \zeta}\left(\partial ^i_b\partial
^j_b-\delta^{ij}\vec{\partial}_b^2/2\right)\delta
q_T'(x,b,\mu,x\zeta,\rho)\nonumber\\&&=\left(K(b,\mu,\rho)+G(x\zeta,\mu,\rho)\right)\left(\partial
^i_b\partial ^j_b-\delta^{ij}\vec{\partial}_b^2/2\right)\delta
q_T'(x,b,\mu,x\zeta,\rho) \ .
\end{eqnarray}

As a final remark, we would like to point out that the energy
evolution equations for the TMD quark distributions in the
Drell-Yan process will be the same as the above for the DIS
process. This is because we have the universality for the parton
distributions between the two processes \cite{Col02,BelJiYua03}.

\section{Resummation For the Structure Functions in Polarized Semi-Inclusive DIS}

In the physical cross sections with two widely separated scales,
say, $P_\perp$ and $Q$, there are large double logarithms of the
type $\alpha_s^n\ln^{2n} Q^2/P_\perp^2$ as well as sub-leading
ones. To have a reliable theoretical prediction, one has to re-sum
these contributions. In this section, we perform the resummation
for the large logarithms in polarized semi-inclusive DIS by
solving the Collins-Soper evolution equations obtained in the last
section.

In \cite{JiMaYu04p}, we have obtained the factorization formulas
for the various structure functions. In the impact parameter $b$
space, they can be expressed as \cite{JiMaYu04p},
\begin{eqnarray}
\tilde F_{UU}^{(1)}(b)&=&q(x_B,z_hb)\hat
q(z_h,b)S^+(b)H_{UU}^{(1)}(Q^2)\
, \nonumber\\
\tilde F_{LL}(b)&=&\Delta q_L(x_B,z_hb)\hat
q(z_h,b)S^+(b)H_{LL}(Q^2)\
, \nonumber\\
\tilde F_{UT}^{(1)}(b)&=&\frac{-1}{M
z_h}\partial_b^i\left[\left(\partial_b^iq_T(x_B,z_hb)\right)\hat
q(z_h,b)S^+(b)H_{UT}^{(1)}(Q^2)\right] \ ,\nonumber\\
\tilde F_{UT}^{(2)}(b)&=&\frac{-1}{M_h}\partial_b^i\left[\delta
q_T(x_B,z_hb)\left(\partial_b^i\delta\hat q(z_h,b)\right)S^+(b)H_{UT}^{(2)}(Q^2)\right]\ ,\nonumber\\
\tilde F_{LT}(b)&=&\frac{-1}{M
z_h}\partial_b^i\left[\left(\partial_b^i\Delta
q_T(x_B,z_hb)\right)\hat
q(z_h,b)S^+(b)H_{LT}(Q^2)\right]\ ,\nonumber\\
\tilde
F_{UU}^{(2)}(b)&=&\frac{\partial_b^i\partial_b^j-\vec{\partial}_b^2\delta^{ij}}{MM_h
z_h}\left[\left(\partial_b^i\delta
q(x_B,z_hb)\right)\left(\partial_b^j\delta \hat
q(z_h,b)\right)S^+(b)H_{UU}^{(2)}(Q^2)\right] \ ,\nonumber\\
\tilde
F_{UL}(b)&=&\frac{\partial_b^i\partial_b^j-\vec{\partial}_b^2\delta^{ij}}{MM_h
z_h}\left[\left(\partial_b^i\delta
q_L(x_B,z_hb)\right)\left(\partial_b^j\delta\hat
q(z_h,b)\right)S^+(b)H_{UL}(Q^2)\right] \ ,\nonumber\\
\tilde
F_{UT}^{(3)}(b)&=&\frac{-4\partial_b^i\partial_b^j\partial_b^k+2\delta^{ik}\vec{\partial}_b^2\partial^j_b}{M^2M_h
z_h^2}\left[\left((\partial_b^i\partial_b^j-\delta^{ij}\vec{\partial}_b^2/2)\delta
q_T'(x_B,z_hb)\right)\right. \nonumber \\
&& \times \left. \left(\partial_b^k\delta\hat
q(z_h,b)\right)S^+(b)H_{UT}^{(3)}(Q^2)\right] \ , \label{sfb}
\end{eqnarray}
where we followed the notations used in \cite{JiMaYu04p}, and the
parton distributions and fragmentation functions are calculated at
the energy scale $x_B^2\zeta^2=\hat\zeta^2/z_h^2=\rho Q^2$.
$\tilde F(b)$ are the Fourier transformation of the structure
functions in the impact parameter space. For the unpolarized
structure function, we define
\begin{equation}
\tilde F_{UU}^{(1)}(b)=\int d^2 P_{h\perp}
F_{UU}^{(1)}(P_{h\perp}) e^{i\vec{P}_{h\perp}\cdot \vec b_\perp} \
,
\end{equation}
and similarly for $F_{LL}$. Others denote
transverse-momentum-weighted Fourier transformations, for example,
\begin{equation}
\tilde F_{UT}^{(1)}(b)=\int d^2 P_{h\perp}
|\vec{P}_{h\perp}|F_{UT}^{(1)}(P_{h\perp})
e^{i\vec{P}_{h\perp}\cdot \vec b_\perp} \ ,
\end{equation}
and similarly for $\tilde F_{UT}^{(2)}$ and $\tilde F_{LT}$. For
$\tilde F_{UU}^{(2)}$ and $F_{UL}$, we define
\begin{equation}
\tilde F_{UU}^{(2)}(b)=\int d^2 P_{h\perp}
|\vec{P}_{h\perp}|^2F_{UU}^{(2)}(P_{h\perp})
e^{i\vec{P}_{h\perp}\cdot \vec b_\perp} \ .
\end{equation}
For $\tilde F_{UT}^{(3)}$, we define
\begin{equation}
\tilde F_{UT}^{(3)}(b)=\int d^2 P_{h\perp}
|\vec{P}_{h\perp}|^3F_{UT}^{(3)}(P_{h\perp})
e^{i\vec{P}_{h\perp}\cdot \vec b_\perp} \ .
\end{equation}

In \cite{JiMaYu04}, the large logarithms have been re-summed for
the unpolarized structure function by solving the relevant
Collins-Soper equation for the TMD quark distribution and
fragmentation function. From Eq.~(\ref{sfb}) and the results of
the previous section concerning the Collins-Soper evolution for
the polarized quark distributions, we conclude that the polarized
structure functions have the same evolution equation as the
unpolarized case. In the following, we take the structure function
$F_{UT}^{(1)}$, depending on the transversely-polarized nucleon
spin, as an example to demonstrate the re-summation procedure.

Rewriting the structure function $F_{UT}^{(1)}$ as,
\begin{equation}
\tilde F_{UT}^{(1)}(b,Q^2)=\frac{-1}{M z_h}\partial_b^i {\cal
F}_{UT}(b,Q^2) \ , \label{fut}
\end{equation}
where
\begin{equation}
{\cal F}_{UT}(b,Q^2)=\left(\partial_b^iq_T(x_B,z_hb)\right)\hat
q(z_h,b)S^+(b)H_{UT}^{(1)}(Q^2) \ .
\end{equation}
Since the Sivers function $q_T$ and the unpolarized fragmentation
function $\hat q$ obey the same Collins-Soper evolution equation
as that for the unpolarized quark distribution, we have the
following derivative equation respect to $Q^2$
\cite{ColSopSte85,JiMaYu04},
\begin{equation}
Q^2\frac{\partial}{\partial Q^2}{\cal
F}_{UT}(x_B,z_h,b,Q^2)=\left[K(b\mu,g(\mu),\rho)+G_{UT}'(Q/\mu,g(\mu),\rho)\right]{\cal
F}_{UT}(x_B,z_h,b,Q^2) \ ,
\end{equation}
where $K$ is the same as before, and $G_{UT}'$ contains additional
contribution from the hard part. The sum of $K$ and $G_{UT}'$ has
no dependence on $\rho$ although they separately might have. The
solution to this differential equation has the following form
\cite{ColSopSte85},
\begin{eqnarray}
{\cal F}_{UT} (x_B,z_h,b,Q^2)&=&{\cal
F}_{UT}(x_B,z_h,b,\mu_1^2/C_2^2)e^{-S(Q^2,b,C_2)}\ ,
\label{cssres}
\end{eqnarray}
where the distribution and fragmentation function are evaluated at
$x_B^2\zeta^2=\hat\zeta^2/z_h^2=\rho\mu_L^2/C_2^2$. The Sudakov
suppression form factor reads,
\begin{eqnarray}
{\cal
S}(Q^2,\mu_L^2,b,C_2)=\int_{\mu_L}^{C_2Q}\frac{d\bar\mu}{\bar\mu}
\left[\ln \left(\frac{C_2Q^2}{\bar\mu^2}\right)A(b\mu_L,\bar
\mu,\rho)+B(C_2,b\mu_L,\bar \mu,\rho)\right] \ . \label{sud}
\end{eqnarray}
Here $C_2$ is a parameter in the order of $1$, and $\mu_L$ is a
low-energy, but still perturbative, scale. The functions $A$ and
$B$ have perturbative expansions in $\alpha_s$, $A=\sum_n
A^{(n)}(\alpha_s/\pi)^n$ and $B=\sum_n B^{(n)}(\alpha_s/\pi)^n$.
They are defined as
\begin{eqnarray}
A(b\mu_L,\bar \mu,\rho)&=&\gamma_K(\bar
\mu)+\beta\frac{\partial}{\partial
g}K(b\mu_L,g(\bar\mu),\rho)\ , \nonumber\\
B(C_2,b\mu_L,\bar\mu,\rho)&=&-2K(b\mu_L,g(\bar\mu),\rho)-2G_{UT}'(1/C_2,g(\bar\mu),\rho)
\ .
\end{eqnarray}
The $A$-function is the same as that for the unpolarized case
\cite{ColSopSte85,JiMaYu04} since it comes only from the
spin-independent soft part. On the other hand, the $B$-function
contains contribution from the unknown hard part $H_{UT}^{(1)}$,
and hence it could be different from that of the unpolarized one.

Substituting the above results into Eq. (\ref{fut}), we will get
\cite{JiMaYu04}
\begin{eqnarray}
\tilde F_{UT}^{(1)}(b,Q^2)=\frac{-1}{M
z_h}\partial_b^i\left[\left(\partial_b^iq_T(x_B,z_hb)\right)\hat
q(z_h,b)S^+(b)H_{UT}^{(1)}(\mu_L^2/C_2^2)e^{-{\cal
S}(Q^2,\mu_L^2,b)}\right]
 \ .
\end{eqnarray}
From the above derivations, we confirmed that the $\rho$
dependence in the soft factor $K$ does not affect the resummation
of the large logarithms. This is because the resummation concerns
the logarithms of the form $\ln^2Q^2 b^2$ in impact parameter
space, while $\rho$ is just a parameter separating the hard and
soft physics. In addition, there is no $\rho$ dependence in the
Sudakov suppression form factor ${\cal S}$.

The re-summation for all the other structure functions can be done
in a similar way. With these re-summation formulas, one can
further study the $Q^2$ dependence of the $P_\perp$ spectrum of
the polarized cross section and asymmetries in the semi-inclusive
DIS \cite{Boe01}.

If $Q^2$ is not too large, for example, at the order of tens
GeV$^2$, it is legitimate to re-sum only the leading double
logarithms (DL) \cite{ColSopSte85,{DokDyaTro80},{ParPet79}}. In
this approximation, we only need to take into account the first
term in the expansion of $A$ function, neglecting the contribution
from $B$. Since $A^{(1)}=4/3$ from the cusp anomalous dimension
$\gamma_K$, the Sudakov suppression factor reduces to
\begin{equation}
{\cal S}^{(DL)}(Q^2,\mu_L^2,C_2)=\frac{4}{3}\int_{\mu_L}^Q
\frac{d\bar\mu}{\bar\mu} \ln
\left(\frac{C_2Q^2}{\bar\mu^2}\right)\ ,
\end{equation}
in the DL approximation. It only depends on $Q^2$ and $\mu_L^2$,
but not on $b$. Moreover, since the the cusp anomalous dimension
is the same for all the leading-twist TMD quark distribution, the
Sudakov suppression factor will be the same for all the
leading-twist polarized structure functions in (\ref{sfb}). One of
the consequences is that one can predict the $P_{h\perp}$ spectrum
at higher $Q^2$ for semi-inclusive hadron production in DIS from
the result at lower $Q^2$. For example, in the DL approximation,
the unpolarized structure function $F_{UT}$ and the single
transversely polarized structure function $F_{UT}^{(1)}$ have the
following dependence on $Q^2$,
\begin{eqnarray}
F_{UU}(x_B,z_h,P_{h\perp},Q^2)&=&F_{UU}(x_B,z_h,P_{h\perp},\mu_L^2)e^{-{\cal
S}^{(DL)}(Q^2,\mu_L^2)}\nonumber\\
F_{UT}^{(1)}(x_B,z_h,P_{h\perp},Q^2)&=&F_{UT}^{(1)}(x_B,z_h,P_{h\perp},\mu_L^2)e^{-{\cal
S}^{(DL)}(Q^2,\mu_L^2)}\ .
\end{eqnarray}
The polarization asymmetry (the ratio of these two structure
functions) as a function of $P_{h\perp}$ will remain the same for
different $Q^2$ at fixed $x_B$ and $z_h$. This constancy in $Q^2$
has been seen from the comparison of the HERMES data at HERA with
that of CLAS at JLab on various spin asymmetries, with the average
$Q^2$ varying by a factor of three \cite{Harut}. The above
analysis applies to all the leading-twist polarized structure
functions and polarization asymmetries. It will be interested to
test this prediction based on DL approximation with future DIS
experiments at different $Q^2$.

We notice that the above formalism also applies for the Drell-Yan
process, which have plenty data at low transverse momentum and not
very high $Q^2$ \cite{E288}. It will be useful to compare the
above DL approximation prediction with these experimental data,
and gain insight for the transverse momentum dependence for the
TMD quark distributions. We will carry this out in a future
publication. This approach is different from what has been done so
far in the literature
\cite{DavSti84,{ArnKau91},{LadYua94},LanBroLadYua01,{QiuZha01}},
where the predictions solely depend on the integrated parton
distributions and the Collins-Soper-Sterman formalism \cite{ColSopSte85}, 
and one needs nonperturbative
parametrization for the large $b$ behavior of the distributions
and the evolution as well. In our approach, the TMD parton
distributions are important nonperturbative ingredients.

If $Q^2$ is very large (e.g., for $W$ and $Z$ bosons production ),
the above approximation breaks down. One has to take into account
sub-leading logarithmic contributions, perhaps up to $A^{(2)}$ and
$B^{(2)}$ in the expansion of the functions $A$ and $B$
\cite{ColSopSte85,DavSti84,{ArnKau91},{LadYua94},LanBroLadYua01,{QiuZha01},KulSteVog02}.

\section{Conclusions}

In this paper, we have studied the Collins-Soper energy evolution
equation for all the leading-twist TMD quark distributions. The
evolution equation has contributions from both the hard and soft
regions of the gluon momentum. Since both parts are independent of
the quark helicity, the evolution kernel is spin independent.
Based on the evolution equation, we can perform re-summation for
the large double logarithms in the polarized structure functions.

We thank H. Avakian for the communications about the transverse
momentum spectrum of the polarized asymmetries. A.I., X. J. and F.
Y. were supported by the U. S. Department of Energy via grant
DE-FG02-93ER-40762. J.P.M. was supported by National Natural
Science Foundation of P.R. China through grand No.19925520. X. J.
is also supported by an Overseas Outstanding Young Chinese
Scientist grant from NSFC.

\end{document}